\documentclass[10pt,journal,compsoc]{IEEEtran}
\ifCLASSOPTIONcompsoc
   \usepackage[nocompress]{cite}
\else
\fi

\ifCLASSINFOpdf
  \usepackage[pdftex]{graphicx}
\else
  \usepackage[dvips]{graphicx}
\fi
\usepackage[cmex10]{amsmath}
\usepackage{amsfonts}

\usepackage{algpseudocode}
\usepackage{algorithm}
\usepackage{amsmath}%
\usepackage{amsfonts}%
\usepackage{amssymb}%
\newtheorem{theorem}{Theorem}
\newtheorem{principle}[theorem]{Principle}
\newenvironment{proof}[1][Proof]{\textbf{#1.} }{\ \rule{0.5em}{0.5em}}

\usepackage{url}

\newtheorem{example}{Example}
\newtheorem{definition}{Definition}

\usepackage{color, soul}

\begin{document}
%
\title{Resolving Multi-party Privacy Conflicts in Social Media}

\author{Jose M. Such,
Natalia Criado
\IEEEcompsocitemizethanks{\IEEEcompsocthanksitem Jose M. Such is with Security Lancaster, School of Computing and Communications,
Lancaster University, UK. \protect\\
e-mail: j.such@lancaster.ac.uk
\IEEEcompsocthanksitem Natalia Criado is with King's College London, UK. \protect\\
e-mail: ncriado.academia@gmail.com}
\thanks{}}



\IEEEcompsoctitleabstractindextext{
\begin{abstract}
Items shared through Social Media may affect more than one user's privacy --- e.g., photos that depict multiple users, comments that mention multiple users, events in which multiple users are invited, etc. The lack of multi-party privacy management support in current mainstream Social Media infrastructures makes users unable to appropriately control to whom these items are actually shared or not. Computational mechanisms that are able to merge the privacy preferences of multiple users into a single policy for an item can help solve this problem. However, merging multiple users' privacy preferences is not an easy task, because privacy preferences may conflict, so methods to resolve conflicts are needed. Moreover, these methods need to consider how users' would actually reach an agreement about a solution to the conflict in order to propose solutions that can be acceptable by all of the users affected by the item to be shared. Current approaches are either too demanding or only consider fixed ways of aggregating privacy preferences. In this paper, we propose the first computational mechanism to resolve conflicts for multi-party privacy management in Social Media that is able to adapt to different situations by modelling the concessions that users make to reach a solution to the conflicts. We also present results of a user study in which our proposed mechanism outperformed other existing approaches in terms of how many times each approach matched users' behaviour. 
\end{abstract}

}

\maketitle

\IEEEdisplaynotcompsoctitleabstractindextext

\IEEEpeerreviewmaketitle

\section{Introduction}

\IEEEPARstart{H}{undreds} of billions of items that are uploaded to Social Media are co-owned by multiple users \cite{photos}, yet only the user that uploads the item is allowed to set its privacy settings (i.e., who can access the item). This is a massive and serious problem as users' privacy preferences for co-owned items usually conflict, so applying the preferences of only one party risks such items being shared with undesired recipients, which can lead to privacy violations with severe consequences (e.g., users losing their jobs, being cyberstalked, etc.) \cite{thomas2010unfriendly}. Examples of items include photos that depict multiple people, comments that mention multiple users, events in which multiple users are invited, etc. Multi-party privacy management is, therefore, of crucial importance for users to appropriately preserve their privacy in Social Media. 

There is recent evidence that users very often negotiate collaboratively to achieve an agreement on privacy settings for co-owned information in Social Media  \cite{lampinen2011we,wisniewski2012fighting}. In particular, users are known to be generally open to accommodate other users' preferences, and they are willing to make some concessions to reach an agreement depending on the specific situation \cite{wisniewski2012fighting}. However, current Social Media privacy controls solve this kind of situations by only applying the sharing preferences of the party that uploads the item, so users are forced to negotiate \emph{manually} using other means 
such as e-mail, SMSs, phone calls, etc. \cite{besmer2010moving} --- e.g., Alice and Bob may exchange some e-mails to discuss whether or not they actually share their photo with Charlie. The problem with this is that negotiating \emph{manually} all the conflicts that appear in the everyday life may be time-consuming 
because of the high number of possible shared items and the high number of possible accessors (or targets) to be considered by users \cite{thomas2010unfriendly}; e.g., a single average user in Facebook has more than 140 friends and uploads more than 22 photos \cite{fb}. 

Computational mechanisms that can automate the negotiation process have been identified as one of the biggest gaps in privacy management in social media \cite{besmer2010moving,lampinen2011we,wisniewski2012fighting,such2014survey,fogues2015open}. 
The \emph{main challenge} is to propose solutions that can be accepted most of the time by all the users involved in an item (e.g., all users depicted in a photo), so that users are forced to negotiate manually as little as possible, thus minimising the burden on the user to resolve multi-party privacy conflicts.

Very recent related literature proposed mechanisms to resolve multi-party privacy conflicts in social media \cite{wishart2010collaborative,squicciarini2009collective,carminati2011collaborative,hu2012detecting,hu2012multiparty,thomas2010unfriendly}. Some of them \cite{wishart2010collaborative,squicciarini2009collective} need too much human intervention during the conflict resolution process, by requiring users to solve the conflicts \emph{manually} or close to \emph{manually}; e.g., participating in difficult-to-comprehend auctions for each and every co-owned item. 
Other approaches to resolve multi-party privacy conflicts are more automated \cite{thomas2010unfriendly,carminati2011collaborative,hu2012detecting}, but they only consider one fixed way of aggregating user's privacy preferences (e.g., veto voting \cite{thomas2010unfriendly}) without considering how users would actually achieve compromise and the concessions they might be willing to make to achieve it depending on the specific situation. Only \cite{hu2012multiparty} considers more than one way of aggregating users' privacy preferences, but the user that uploads the item chooses the aggregation method to be applied, which becomes a unilateral decision without considering the preferences of the others. 

In this paper, we present the first computational mechanism for social media that, given the individual privacy preferences of each user involved in an item, is able to find and resolve conflicts by applying a different conflict resolution method based on the concessions users' may be willing to make in different situations. 
We also present a user study comparing our computational mechanism of conflict resolution and other previous approaches to what users would do themselves manually in a number of situations. The results obtained suggest our proposed mechanism significantly outperformed other previously proposed approaches in terms of the number of times it matched participants' behaviour in the study.


\section{Background}
\label{sec:back}

Assume a finite set of users $U$, where a finite subset of \emph{negotiating} users $N\subseteq U$, negotiate whether they should grant a finite subset of \emph{target} users\footnote{We defined the set of target users as a subset of the users to remain as general as possible; i.e., without forcing it to satisfy a particular property. However, the set of target users could be further qualified as a particular subset of users satisfying any property without changing the subsequent formalisation; e.g., the set of target users could be defined as the union of all of the negotiating users' online friends.} $T\subseteq U$ access to a particular co-owned item. For instance, Alice and Bob (negotiating users) negotiate about whether they should grant Charlie (target user) access to a photo of them depicted together. For simplicity and without loss of generality, we will consider a negotiation for one item over the course of this paper --- e.g., a photo that depicts the negotiating users together --- and hence, we do not include any additional notation for the item in question.

\vspace{-8pt}
\subsection{Individual Privacy Preferences}

Negotiating users have their own individual privacy preferences about the item --- i.e., to whom of their online friends they would like to share the item if they were to decide it unilaterally. 
In this paper, we assume negotiating users specify their individual privacy preferences using group-based access control, which is nowadays mainstream in Social Media (e.g., Facebook lists or Google+ circles), to highlight the practical applicability of our proposed approach. However, other access control approaches for Social Media could also be used in conjunction with our proposed mechanism --- e.g., relationship-based access control \cite{carminati2009enforcing,fong2011relationship,such2012disclosure} as already shown in \cite{such2014adaptive}, or (semi-)automated approaches like \cite{fang2010privacy,squicciarini2011a3p,danezis2009inferring}. 
Note also that our approach does not necessarily need users to specify their individual privacy preferences for each and every item separately, they could also specify the same preferences for collections or categories of items for convenience according to the access control model being used ---e.g., Facebook users can specify preferences for a whole photo album at once.

Mainstream Social Media (Facebook, Google+, etc.) have predefined groups and also allow users to define their own groups, each of which is composed of a set of friends. Access to items (photos, etc.) can be granted/denied to groups, individuals or both (e.g., all Friends have access to a photo except Charlie). We formally define a group $G\subseteq U$ as a set of users, and the set of all groups defined by a particular user $u$ as $\mathcal{G}_u=\{G_1,\mathellipsis,G_l\}$, so that 
$\bigcap_{G\in\mathcal{G}_u} G = \emptyset$. For instance, Alice may have defined the following groups $\mathcal{G}_{Alice}=\{\texttt{CloseFriends},\texttt{Family},\texttt{Coworkers}\}$ 
to organise her online friends. 

\begin{definition}
A privacy policy $P$ is a tuple $P=\langle A, E\rangle$, where $A$ is the set of groups granted access and $E\subseteq U$ is a set of individual user exceptions.
\end{definition}

The semantics of a group-based privacy policy in most Social Media are: $P.A$ are the groups that are \emph{authorised} (or granted) access to the item; and $P.E$ are a set of individual exceptions --- either users in the authorised groups who are denied access individually or users who are granted access individually because they are in the unauthorised groups (groups not explicitly granted access). Continuing the example above, Alice defines her individual privacy policy for an item as $P_{Alice}=\langle \{\texttt{CloseFriends}\}, \{\texttt{Charlie}\} \rangle$, i.e., Alice wants to share the item only with CloseFriends but excluding Charlie. 
  
\vspace{-8pt}
\subsection{Problem Statement}
Given a set of negotiating users $N=\{n_1,\mathellipsis,n_k\}$ who co-own an item --- i.e., there is one $uploader\in N$ who uploads the item to social media and the rest in $N$ are users affected by the item; and their individual (possibly conflicting) privacy policies $P_{n_1},\mathellipsis,P_{n_k}$
for that item; how can the negotiating users agree on with whom, from the set of the target users $T=\{t_1,\mathellipsis,t_m\}$, the item should be shared?

This problem can be decomposed into:
\begin{enumerate}
\itemsep0cm
\item Given the set of individual privacy policies $P_{n_1},\mathellipsis,P_{n_k}$ of each negotiating user for the item, how can we identify if at least two policies have contradictory decisions --- or \emph{conflicts} --- about whether or not granting target users $T$ access to the item.
\item If conflicts are detected, how can we propose a solution to the conflicts found that respects as much as possible the preferences of negotiating users $N$.
\end{enumerate}

\section{Mechanism Overview}
\label{sec:overview}

\begin{figure}[!h]
\centering
\vspace{-8pt}
\includegraphics[scale=0.4]{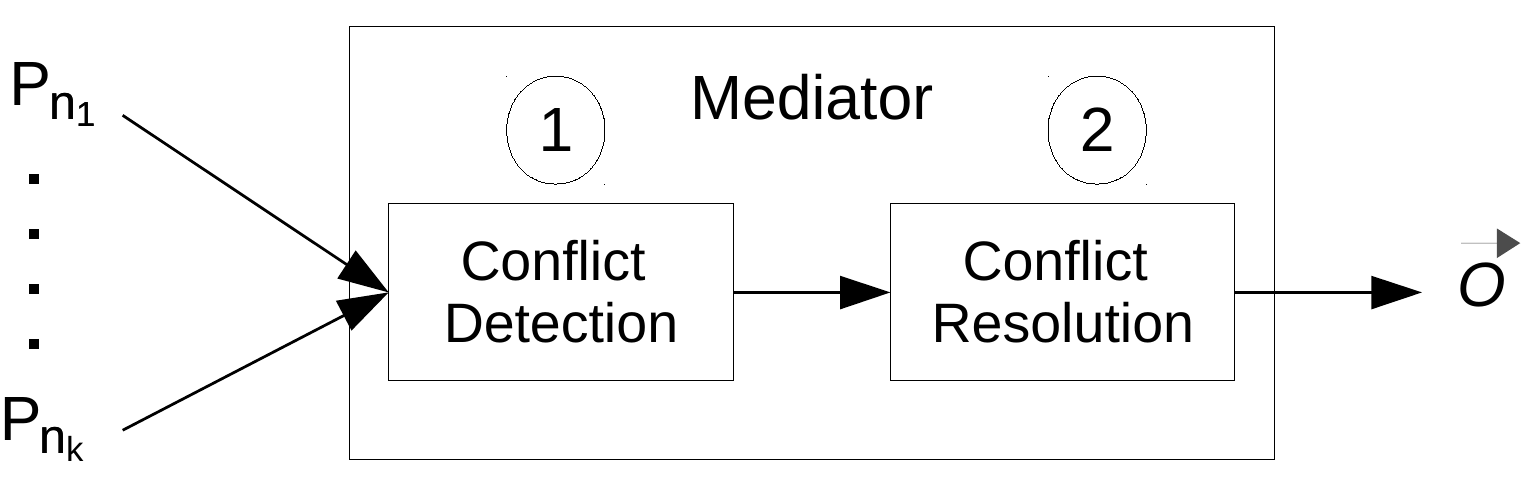}
\vspace{-3pt}
\caption{Mechanism Overview}
\label{fig:overview}
\end{figure}

We propose the use of a mediator that detects conflicts and suggests a possible solution to them. For instance, in most Social Media infrastructures, such as Facebook, Twitter, Google+ and the like, this mediator could be integrated as the back-end of Social Media privacy controls' interface; or it could be implemented as a Social Media application --- such as a Facebook app --- that works as an interface to the privacy controls of the underlying Social Media infrastructure. 
%
%
%
Figure \ref{fig:overview} depicts an overview of the mechanism proposed. 
In a nutshell, the process the mediator follows is:
\begin{enumerate}
\itemsep0cm
\item The mediator inspects the individual privacy policies of all users for the item and flags all the conflicts found (as described in Section \ref{sec:conflicts}). Basically, it looks at whether individual privacy policies suggest contradictory access control decisions for the same target user. If conflicts are found the item is not shared preventively.
\item The mediator proposes a solution for each conflict found. To this aim, the mediator estimates (as described in Section \ref{sec:resolution}) how willing each negotiating user may be to concede by considering: her individual privacy preferences, how sensitive the particular item is for her, and the relative importance of the conflicting target users for her.
\end{enumerate}

If all users accept the solution proposed, it will be applied. Otherwise, users will need to turn into a \emph{manual} negotiation by other means. Note that different approaches could be used to communicate the suggested solutions to users and getting back their feedback, as discussed in Section 7.

\section{Conflict Detection}
\label{sec:conflicts}

We need a way to compare the individual privacy preferences of each negotiating user in order to detect conflicts among them. However, each user is likely to have defined different groups of users, so privacy policies from different users may not be directly comparable. To compare privacy policies from different negotiating users for the same item, we consider the \emph{effects} that each particular privacy policy has on the set of target users $T$. 
%
%
%
Privacy policies dictate a particular action to be performed when a user in $T$ tries to access the item. In particular, we assume that the available actions are either $0$ (denying access) or $1$ (granting access). The action to perform according to a given privacy policy is determined as follows\footnote{Note that the definition of this function will vary according to the access control model used, but it will be defined in a similar way. That is, the idea is to be able to know, given a target user $t$, whether the privacy policy will grant/deny $t$ access to the item regardless of the access control model being used. 
}:


\begin{definition}
Given an user $n\in N$, her groups $\mathcal{G}_n$, her individual privacy policy $P_n=\langle A, E \rangle$, and a user $t\in T$; we define the action function as:
\begin{displaymath}
act(P_n,t) =  
\begin{cases}
1 & \text{if } \exists G\in \mathcal{G}_n: t\in G \wedge G \in P_n.A \wedge t\notin P_n.E\\
1 & \text{if } \exists G\in \mathcal{G}_n: t\in G \wedge G \notin P_n.A \wedge t\in P_n.E\\
0 & \text{otherwise}
\end{cases}
\end{displaymath}
\end{definition}


We also consider so-called \emph{action vectors} $\vec{v}\in\{0,1\}^{\mid T\mid}$; i.e., complete assignments of actions to all users in $T$, such that $v[t]$ denotes the action for user $t\in T$. When a privacy policy is applied to the set of users $T$, it produces such an action vector
%
%
%
%
%
%
, where $v[t]=act(P,t)$.  


If all the action vectors of all negotiating users assign the same action for all target users, then there is no conflict. Otherwise, there are at least two action vectors that assign different actions to the same target user, and there is a conflict. In other words, a conflict arises when some negotiating users would like to grant access to one target user while the others would not. Formally:

\begin{definition}[conflict]
Given a set of negotiating users $N$ and a set of target users $T$; a target user $t\in T$ is said to be \emph{in conflict} iff $\exists a,b\in N$ with individual privacy policies $P_a$ and $P_b$ respectively, so that $v_a[t]\neq v_b[t]$.


\end{definition}

Further, we say that the set of \emph{users in conflict} $C\subseteq T$, is the set that contains all the target users that are in conflict. 

\begin{algorithm}
\begin{footnotesize}
\begin{algorithmic}[1]
\Require $N$, $P_{n_1},\mathellipsis,P_{n_{\mid N\mid}}$, $T$  
\Ensure $C$
\ForAll{$n\in N$}
\ForAll{$t\in T$}
\State $v_n[t]\gets 0$
\ForAll{$G\in P_n.A$}
\If {$\exists u\in G, u=t$}
\State $v_n[t]\gets 1$
\EndIf
\EndFor
\EndFor
\ForAll{$e\in P_n.E$}
\State $v_n[e]\gets \neg v_n[e]$
\EndFor
\EndFor
\State $C\gets \emptyset$
\ForAll{$t\in T$} 
\State Take $a\in N$
\ForAll {$ b \in N \setminus \{a\}$ }
\If{$v_a[t] \neq v_b[t]$ }
    \State $C\gets C\cup \{t\}$
\EndIf
\EndFor
\EndFor
\end{algorithmic}
\end{footnotesize}
\caption{Conflict Detection}
\label{al:cd}
\end{algorithm}

%
%


The mediator runs Algorithm \ref{al:cd} to detect conflicts by harvesting the {\em users in conflict} set $C$. The complexity of the algorithm is \emph{polynomial} and it mainly depends on the number of negotiating users, target users, groups granted access, and users in each group granted access. In the worst case, the complexity is $\mathcal{O}(|U|^3)$, when all users $U$ are negotiators and targets; all groups of all negotiators are granted access; and, for each negotiator, there are as many groups as users or all users are in one group\footnote{Recall groups are disjoint. Otherwise, the complexity is $\mathcal{O}(|U|^4)$.}. If Algorithm \ref{al:cd} does not detect any conflict --- i.e., $C=\emptyset$, it will return to the users without changes to their preferred privacy policies. If Algorithm \ref{al:cd} detects conflicts, the mediator will then run the conflict resolution module, which is described in the following section.


\begin{example}
Assume a set of users $U=\{\texttt{Alice},\texttt{Bob},$ $ \texttt{Charlie}, \texttt{Dan}, \texttt{Eve}, \texttt{Frank}\}$. 
%
Negotiating users $N=\{\texttt{Alice}, \texttt{Bob}\}$ are in the process of deciding to which target users $T=\{\texttt{Charlie}, \texttt{Dan}, \texttt{Eve}, \texttt{Frank}\}$ they grant access to a photo in which both of them are depicted. Negotiating users defined the following groups: Alice defined $\mathcal{G}_{\texttt{Alice}}=\{\texttt{MyFriends}\}$ so that $\texttt{MyFriends}=\{\texttt{Charlie}, \texttt{Dan}, \texttt{Eve}\}$; and Bob defined $\mathcal{G}_{\texttt{Bob}}=\{\texttt{CloseFriends},\texttt{Family}\}$ so that $\texttt{CloseFriends}=\{\texttt{Charlie},\texttt{Eve}\}$ and $\texttt{Family}=\{\texttt{Dan},\texttt{Frank}\}$. Now, assume that negotiating users have the following individual privacy policies for the photo: Alice has $P_{\texttt{Alice}}=\langle\{\texttt{MyFriends}\},\{\texttt{Eve}\}\rangle$ so that $\vec{v}_{\texttt{Alice}}=(1,1,0,0)$ --- i.e., only Charlie and Dan would be granted access to the photo; and Bob has $P_{\texttt{Bob}}=\langle\{\texttt{CloseFriends},\texttt{Family}\},\emptyset\rangle$ so that $\vec{v}_{\texttt{Bob}}=(1,1,1,1)$ --- i.e., all target users Charlie, Dan, Eve and Frank would be granted access to the photo. As $v_{\texttt{Alice}}[\texttt{Eve}]\neq v_{\texttt{Bob}}[\texttt{Eve}]$ and $v_{\texttt{Alice}}[\texttt{Frank}]\neq v_{\texttt{Bob}}[\texttt{Frank}]$, the set of users in conflict is $C=\{\texttt{Eve},\texttt{Frank}\}$.
\end{example}

\section{Conflict Resolution}
\label{sec:resolution}
When conflicts are detected, the mediator suggests a solution according to 
the following principles:

\begin{itemize}
\item \textbf{Principle 1:} An item should not be shared if it is detrimental to one of the users involved --- i.e., users refrain from sharing particular items because of potential privacy breaches \cite{sleeper2013post} and other users allow that as they do not want to cause any deliberate harm to others \cite{lampinen2011we,besmer2010moving}.
\item \textbf{Principle 2:} If an item is not detrimental to any of the users involved and there is any user for whom sharing is important, the item should be shared --- i.e., users are known to accommodate others' preferences \cite{lampinen2011we,wisniewski2012fighting,besmer2010moving}.
\item \textbf{Principle 3:} For the rest of cases, the solution should be consistent with the majority of all users' individual preferences --- i.e., when users do not mind much about the final output \cite{wisniewski2012fighting,lampinen2011we,besmer2010moving}.
\end{itemize}

We shall now describe the framework to model these principles and Appendix\ref{sec:proofs} shows the proofs that the framework follows the principles above. In a nutshell, the mediator computes a solution to the conflicts as detailed in Section 5.3,  based on the three principles above, which are operationalised as concession rules as detailed in Section 5.2. Concessions rules are in turn instantiated based on the preferred action of each user for the conflict (dictated by each user's individual privacy policy) as well as an \emph{estimated willingness to change} that action (detailed in Section 5.1). 

\subsection{Estimating the Willingness to change an action}
In order to find a solution to the conflict that can be acceptable by all negotiating users, it is key to account for how important is for each negotiating user to grant/deny access to the conflicting target user. In particular, the mediator estimates how \emph{willing} a user would be to change 
the action (granting/denying) she prefers for a target agent in order to solve the conflict 
based on two main factors: the sensitivity of the item and the relative importance of the conflicting target user. 

\subsubsection{\hspace{15pt} Estimating Item Sensitivity} 
If a user feels that an item is very sensitive for her\footnote{Note that we particularly stress that an item is sensitive for someone. This is because the same item may be seen as having different sensitivity by different people.}, she will be less willing to accept sharing it than if the item is not sensitive for her \cite{sleeper2013post,wang2011regretted}. One way of eliciting item sensitivity would be to ask the user directly, but this would increase the burden on the user. Instead, the mediator estimates how sensitive an item is for a user based on how strict is her individual privacy policy for the item \cite{squicciarini2011a3p}, so that the stricter the privacy policy for the item the more sensitive it will be. Intuitively, the lower the number of friends granted access, the stricter the privacy policy, hence, the more sensitive the item is. Moreover, not all friends are the same; i.e., users may feel closer to some friends than others and friends may be in different groups representing different social contexts. Thus, both the group and the strength of each relationship are considered when estimating the strictness of privacy policies and, therefore, the sensitivity of items.

The mediator can use any of the existing tools to automatically obtain relationship strength (or \emph{tie strength}) values for all the user's friends for particular Social Media infrastructures such as Facebook \cite{fogues2013bff,gilbert2009predicting} and Twitter \cite{gilbert2012predicting} with minimal user intervention. Even if the mediator would not be able to use these tools, users could be asked to self-report their tie strength to their friends, which would obviously mean more burden on the users but would still be possible. Whatever the procedure being used, the mediator just assumes that the tie strength value assigned for each pair of friends $a$ and $b$ is given by a function $\tau(a,b)$, so that $\tau: U \times U \rightarrow \{0,\mathellipsis,\delta\}$, where $\delta$ is the maximum positive integer value in the tie strength scale used\footnote{The maximum tie strength value $\delta$ depends on the tool used. For example, in Fogu\'{e}s et al. \cite{fogues2013bff} $\delta=5$; i.e., six levels of tie strength, which would map to, for instance, the friend relationship as: 0-no relationship, 1-acquaintance, 2-distant friend, 3-friend, 4-close friend, 5-best friend.}.

Based on these values, the mediator considers how strict is a user's individual privacy policy as an \emph{estimate of} the sensitivity of an item by calculating the minimum tie strength needed in each group to have access to the item and averaging it across groups. That is, if a privacy policy only grants users with close relationships (i.e., friends with high tie strength values) access to an item, then the item will be estimated as sensitive, since the privacy policy is very strict (i.e., the average minimum tie strength across groups to have access to the item is very high). On the contrary, if a privacy policy grants users with low tie strengths across groups, then the item will be estimated as less sensitive, since the privacy policy is less strict.

\begin{definition}
Given a user $n\in N$, her groups $\mathcal{G}_n$, and her individual privacy policy $P_n$ for an item, the sensitivity of the item for $n$ is estimated as:
\vspace{-5pt}
\begin{displaymath}
\mathcal{S}_n = \frac{1}{\mid \mathcal{G}_n \mid}\sum_{G\in \mathcal{G}_n} \mathcal{T}_n(G)
\end{displaymath}
\end{definition}
\vspace{-5pt}
where $\mathcal{T}_n(G)$ is the strictness of the privacy policy in group $G$, defined as the minimum tie strength needed in group $G$ to have access to the item:
\vspace{-5pt}

\begin{displaymath}
\mathcal{T}_n(G) = \min_{t\in G} f(n,t)
\end{displaymath}

and $f(n,t)$ is based on the tie strength between users $n$ and $t$. However, this function considers differently situations where $t$ is given access and situations where $t$ is denied access. In particular, if user $t$ is granted access, then function $f$ returns the tie strength between users $n$ and $t$. On the contrary, if user $t$ is denied access, then this user must not be considered when determining the policy strictness for the group and function $f$ returns the maximum tie strength value (recall that $\mathcal{T}_n(G)$ is defined as the minimum value returned by function $f$ for all users in a group). More formally, $f(n,t)$ is defined as follows:
\vspace{-10pt}

\begin{displaymath}
f(n,t)=
\begin{cases}
\tau(n,t) & \text{iff } act(P_n,t)=1\\
\delta & \text{iff } act(P_n,t)=0\\
\end{cases} 
\end{displaymath}

\subsubsection{\hspace{15pt} Estimating the relative importance of the conflict}
Now the focus is on the particular conflicting target user --- i.e., the target user for which different negotiating users prefer a different action (denying/granting access to the item). The mediator estimates how \emph{important} a conflicting target user is for a negotiating user by considering both tie strength with the conflicting target user \cite{green06,houghton2010privacy,wiese2011you} and the group (relationship type) the conflicting target user belongs to \cite{kairam2012talking,fang2010privacy,danezis2009inferring}, which are known to play a crucial role for privacy management. 
For instance, Alice may decide she does not want to share a party photo with her mother, who has a very close relationship to Alice (i.e., tie strength between Alice and her mother is high). This signals that not sharing the photo with her mother is very important to Alice, e.g., teens are known to hide from their parents in social media \cite{boyd2011social}. Another example would be a photo in which Alice is depicted together with some friends with a view to a monument that she wants to share with all her friends. If some of her friends that appear in the monument photo also want to include Alice's acquaintances, it is likely she would accept as she already wants to share with all her friends (whether close or distant). Thus, the mediator estimates the relative importance of a particular conflicting user considering both the tie strength with this user in general and within the particular group (relationship type) she belongs to. In particular, the mediator estimates the relative importance a conflicting target user has for a negotiating user as the difference between the tie strength with the conflicting user and the strictness of the policy for the group the conflicting user belongs to. If the conflicting target user does not belong to any group of the negotiator; then the relative importance is estimated considering the item sensitivity instead as there is no group information.
\vspace{5pt}

\begin{definition}
Given a user $n\in N$, her groups $\mathcal{G}_n$, and a conflicting user $c\in C$, the relative importance of $c$ for $n$ is estimated as follows:
\begin{displaymath}
I_n(c) = \begin{cases}
   \mid\mathcal{T}_n( G ) - \tau(n,c)\mid        & \text{if } \exists G\in \mathcal{G}_n: c\in G \\
   \mid S_n - \tau(n,c)\mid    & \text{otherwise}
  \end{cases}
\end{displaymath}
\end{definition}

For instance, assume Alice would like to share with all her friends ---i.e., $\mathcal{T}_{Alice}(Friends)=1$--- but not with Charlie, who is close friend of her ---i.e., $\tau(Alice,Charlie)=5$. The relative importance would be calculated as $I_{Alice}(Charlie) = \mid 1 - 5\mid=4$, which means that the action Alice prefers for Charlie is quite important to her; e.g., Alice could be creating an event in which she invites all her friends except Charlie because the event is a surprise for Charlie's birthday, so sharing with Charlie would mean ruining the surprise party. In the very same way, if Alice would like to share an item only with her best friend ---i.e., $\mathcal{T}_{Alice}(Friends)=5$, the relative importance of denying access to an acquaintance would be high too ---i.e., if Peter is an acquaintance of Alice such that  $\tau(Alice,Peter)=1$, then $I_{Alice}(Peter) = \mid 5 - 1\mid=4$.

\subsubsection{\hspace{15pt}Estimating Willingness}

Finally, the mediator estimates the \emph{willingness} to change 
the preferred action (granting/denying) for a conflicting target user accounting for both the sensitivity of the item and the relative importance of the conflicting target user as detailed above. If both sensitivity and relative importance are the highest possible, then the willingness to change should be minimal. On the contrary, if both sensitivity and relative importance are the lowest possible, then the willingness to change should be maximal. Thus, we define willingness as a distance (in a 2-dimensional space) between the values of both item sensitivity and relative importance and the maximum possible values for both --- as shown above, both measures are defined in tie strength units and have $\delta$ as their maximum value\footnote{The calculations and meaning for sensitivity and relative importance are different and they may render different values for the same conflict, so they are considered as two different dimensions.}
. We chose for this the Canberra distance\footnote{Given two n-dimensional vectors $\vec{p}$ and $\vec{q}$, the Canberra distance \cite{lance1967mixed} is defined as:
\vspace{-16pt}
 \begin{displaymath}
d(\vec{p},\vec{q}) = \sum_{i=1}^{n} \frac{\mid p_i-q_i\mid}{\mid p_i\mid + \mid q_i\mid} 
\end{displaymath}} instead of other distances like Euclidean, Manhattan, or Chebyshev because it is a relative and not absolute distance metric --- so that it would work in the same way regardless of the $\delta$ value being used.

\begin{definition}
Given user $n\in N$, her preferred privacy policy $P_n$, the maximum tie strength value 
 $\delta$, a conflicting target user $c\in C$, the willingness of user $n$ to accept changing her most preferred action for $c$ is a function $\mathcal{W}:N\times C \rightarrow [0,1]$ so that: 
\label{def:utility}
\end{definition}
\begin{displaymath}
\mathcal{W}(n,c) = \frac{1}{2}\cdot\left(\frac{\mid \delta-I_n(c) \mid}{\delta+I_n(c)} + \frac{\mid \delta-S_n \mid}{\delta+S_n}\right)
\end{displaymath}

Note that the only difference from a 2-dimensional Canberra distance is that we divide by 2 the final result to normalise the willingness into a real value within the [0,1] interval for convenience to model concessions as shown in the following section (Section \ref{sec:will}).

\begin{example}
Suppose Example 1 and that we would like to obtain the willingness of Alice and Bob to accept changing their preferred actions for the conflicts found $C=\{Eve,Frank\}$. Suppose also that the tie strength between users are those given in Table \ref{tab:int}. Table \ref{tab:ut} shows all the willingness values for each of the conflicts and possible solutions. For instance, to calculate $\mathcal{W}(\text{Alice},\text{Eve})$, the mediator first calculates the item sensitivity and the relative importance of Eve as follows: 
\vspace{-3pt}
\begin{align*}
\mathcal{S}_\text{Alice} & = \frac{1}{\mid \mathcal{G}_\text{Alice} \mid}\sum_{G\in \mathcal{G}_\text{Alice}} \mathcal{T}_\text{Alice}( G ) = \mathcal{T}_\text{Alice}( \text{MyFriends} ) = 2 
\end{align*}

and 
\vspace{-10pt}

\begin{align*}
I_{\text{Alice}}(\text{Eve}) \! = \! |\mathcal{T}_\text{Alice}( \text{MyFriends} ) \! - \! \tau(\text{Alice},\text{Eve})| \! = \! |2-1| \! = \! 1\\
\end{align*}

\vspace{-5pt}

Therefore, the willingness is:
\vspace{-10pt}

\begin{align*}
\mathcal{W}(\text{Alice},\text{Eve}) & = \frac{1}{2}\cdot\left(\frac{\mid \delta-I_\text{Alice}(\text{Eve}) \mid}{\delta+I_\text{Alice}(\text{Eve})} + \frac{\mid \delta-S_\text{Alice} \mid}{\delta+S_\text{Alice}}\right) \\
& = \frac{1}{2}\cdot\left(\frac{\mid 5-1 \mid}{5+1} + \frac{\mid 5-2 \mid}{5+2}\right)\\
& = \frac{1}{2}\cdot\left(\frac{4}{6} + \frac{3}{7}\right)\approx 0.55
\end{align*}

We can see in Table \ref{tab:ut} that the mediator would estimate Alice's willingness to grant Eve access to the item higher than Alice's willingness to grant Frank access to the item  --- recall Alice's preferred action for both Eve and Frank is to deny access, so the mediator estimates \emph{willingness to grant access}. The reason for the estimated willingness is that, though the item seems not very sensitive for Alice ($S_a=2$), Eve is closer to Alice than Frank, who seems not to be friend of Alice at all or be a very distant acquaintance because of a 0 tie strength. We can also see in Table \ref{tab:ut} that the mediator would estimate Bob's willingness not to share with Eve to be lower than Bob's willingness not to share with Frank --- recall Bob's preferred action for both Eve and Frank is to grant access, so the mediator estimates \emph{willingness to deny access}. This is because Eve seems to have higher relative importance than Frank for Bob; i.e., Eve seems to be best friends with Bob (high tie strength), so it is plausible to believe Bob would definitely want to share with his best friend and would be unwilling to accept not sharing with her. 

\end{example}

\begin{table}[h]
\begin{center}
\begin{tabular}{|c|c|c|c|c|} \hline
&Charlie & Dan & Eve & Frank \\ \hline
Alice  & 4 & 2 & 1 & 0 \\ \hline
Bob  & 3 & 2 & 5 & 2 \\ \hline
\end{tabular}
\caption{Tie strength for Example 2, with $\delta=5$ according to \cite{fogues2013bff}.}
\label{tab:int}
\end{center}
\vspace{-0.7cm}
\end{table}

\begin{table}[h]
\begin{center}
\begin{tabular}{|c|c|c|} \hline
&Eve & Frank \\ \hline
Alice  & 0.55 & 0.43 \\ \hline
Bob  & 0.34 & 0.71\\ \hline
\end{tabular}
\caption{Willingness for Example 2.}
\label{tab:ut}
\end{center}
\vspace{-0.7cm}
\end{table}

\subsection{Modelling Concessions}
\label{sec:will}
\begin{figure}[!t]
\centering
\includegraphics[scale=0.2]{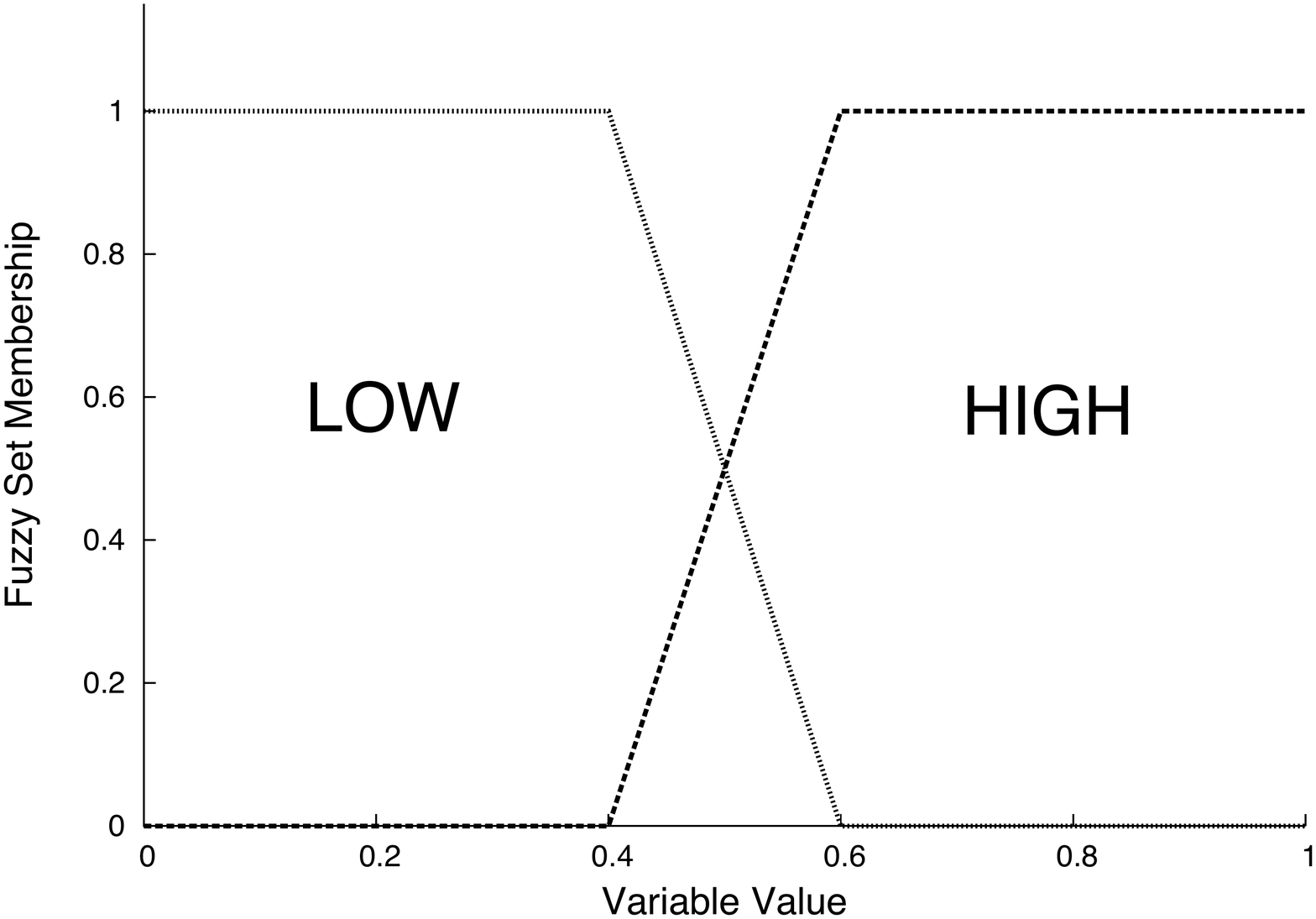}
\vspace{-0.3cm}
\caption{Fuzzy sets low and high.}
\label{fig:fuzzy}
\vspace{-0.3cm}
\end{figure}

As suggested by existing research \cite{lampinen2011we,wisniewski2012fighting,besmer2010moving}, negotiations about privacy in social media are collaborative most of the time. That is, users would consider others' preferences when deciding to whom they share, so users may be willing to concede and change their initial most-preferred option. Being able to model the situations in which these concessions happen is of crucial importance to propose the best solution to the conflicts found --- one that would be acceptable by all the users involved. 
To this aim, the mediator models users' decision-making processes during negotiations based on the willingness to change an action (defined above) as well as on findings about \emph{manual} negotiations in this domain, like the ones described in  \cite{lampinen2011we,wisniewski2012fighting,besmer2010moving}. Users' decision making on continuous variables, like the willingness to change an action, is commonly modelled using fuzzy sets that characterize intervals of the continuous variables \cite{zadeh1992calculus}. Figure \ref{fig:fuzzy} depicts the intervals the mediator considers for the willingness to change an action, which can be 
low or high\footnote{Note that by design, as we are dealing with privacy, we take a conservative approach and the cutting point of exactly 0.5 is considered low.}. Based on this, the following fuzzy IF-THEN rules to model \emph{concessions} in different situations as described below according to the three principles stated above.

\subsubsection*{I do not mind (IDM) rule}

Users are generally willing to accommodate others' sharing preferences \cite{wisniewski2012fighting,lampinen2011we}, so if they do not mind much about which action is finally applied, they will concede and accept applying the action that is not the most preferred for them. In particular, if the willingness to accept the action that is not the preferred one is high, then this may mean that the user would not mind much conceding and accepting that action for the conflicting target user. Assuming a negotiating user $a\in N$, and a conflicting target user $c\in C$, 
this concession can be formalised as the following fuzzy IF-THEN rule:

\vspace{-0.5cm}

\begin{equation}
\tag{IDM}
\texttt{\textbf{IF} $\mathcal{W}(a,c)$ \textbf{IS} high \textbf{THEN} concede}
\label{IDM}
\end{equation}


Note that \texttt{concede} means that user $a$ would accept changing her initial most preferred action to reach an agreement. Thus, users that would initially prefer granting the particular conflicting target user access to the corresponding item would accept denying access, and users that would initially prefer denying the particular conflicting target user access to the corresponding item would accept granting access. For instance, Alice and Bob could be depicted in a photo with very low sensitivity --- e.g., a photo in which both Alice and Bob are depicted with a view to a monument --- and both of them could have defined privacy policies for the photo so that all their friends can see it. Suppose that Charlie is friend of Alice but is distant acquaintance of Bob, so according to Alice's privacy policy Charlie should be granted access to the photo but according to Bob's privacy policy Charlie should not be granted access to the photo. However, given that the photo is not sensitive for Bob, Bob would probably accept sharing also with Charlie and solve the conflict.


\subsubsection*{I understand (IU) rule}

Even when the willingness to change an action is low for some of the negotiating users, users do not want to cause any deliberate harm to their friends and will normally listen to their objections \cite{wisniewski2012fighting}. That is, if the item may be detrimental to some of the negotiating users, so that they prefer denying a conflicting target user access and the willingness to grant access is low, then other users whose most preferred action for the target user is granting access and the willingness to accept denying is also low would \emph{concede} and accept denying access to the conflicting target user. Indeed, considering others self-presentation online has been reported as a way of reaffirming and reciprocating user's relationships \cite{wisniewski2012fighting,green06}. Assuming a negotiating user $a\in N$, and a conflicting target user $c\in C$
, this concession can be formalised as the following fuzzy IF-THEN rule:






\vspace{-0.5cm}

\begin{multline}
\tag{IU}
\texttt{\textbf{IF} } \ \mathcal{W}(a,c) \texttt{ \textbf{IS}  low } \ \wedge \ v_a[c]=1 \ \wedge \\
 \exists b\in N, \ \mathcal{W}(b,c) \texttt{ \textbf{IS}  low } \ \wedge \ v_b[c]=0 \\
 \texttt{ \textbf{THEN} concede}
\label{IU}
\end{multline}



For instance, Alice, Bob, and Charlie are depicted together in a photo in which Bob is clearly inebriated. Initially, Alice and Charlie might very much like to share the photo with friends because Alice, Bob and Charlie could agree they had a very good time together that day in which the photo was taken. However, Alice and Charlie would probably understand the privacy implications this may entail to Bob. Thus, if Bob opposes sharing the photo, Alice and Charlie would probably accept not sharing the photo.


\subsubsection*{No concession (NC) rule}
For the other cases in which neither IDM nor IU applies, then the mediator estimates that a negotiating user would not concede and would prefer to stick to her preferred action for the conflicting target user. For completeness, this can be formalised as the following fuzzy IF-THEN rule assuming a negotiating user $a\in N$, and a conflicting target user $c\in C$:

\vspace{-0.5cm}

\begin{multline}
\tag{NC}
\texttt{\textbf{IF} } \ \mathcal{W}(a,c) \texttt{ \textbf{IS}  low}  \ \wedge \\
( v_a[c]=0 \ \vee  
 (\not\exists b\in N: \mathcal{W}(b,c) \texttt{ \textbf{IS}  low} \ \wedge \ v_b[c]=0)) \\
 \texttt{ \textbf{THEN} do not concede}
\label{NC}
\end{multline}

For instance, when the willingness to accept granting access to the item is low, users very much seek to avoid sharing the item \cite{sleeper2013post}, because it can cause them a privacy breach; i.e., a sensitive item ends up shared with someone they would not like ---e.g., in the example above, Bob would most probably not accept sharing the photo in which he appears inebriated with Alice and Charlie's friends because he might feel embarrassed about the photo and would prefer that no one sees it. 

\subsection{Computing Conflict Resolution}
The mediator computes the solution for each conflict found by applying the concession rules defined above. The solution will be encoded into an action vector $\vec{o}$, so that $o[t]$ contains the action for target user $t$. If $t$ is not conflicting, the mediator assigns to this target user the action shared by all negotiation users. If $t$ is conflicting, the mediator assigns to $o[t]$ its proposal to solve the conflict. To this aim, the mediator executes Algorithm \ref{al:cr}. In particular, for each conflicting target user $t$:

\begin{itemize}
\item If for all negotiating users, their willingness to accept changing their preferred action for the conflicting target user is high
, then, according to concession rule \ref{IDM}, the mediator assumes that all users are willing to concede if need be, so that the final action to be applied for target user $t$ can be both grating and denying. In order to select one of these two actions, the mediator runs a modified majority voting rule (Lines 3-6). In particular, this function selects the action that is most preferred by the majority of users. In case that there is a tie --- i.e., the number of users who prefer granting and the number of users who prefer denying is the same, then the uploader is given an extra vote. Note that this function is only used if all the users have a high willingness to accept the action that is not the most preferred for them. That is, it does not really make much of a difference for them which action is finally taken, and all of them are willing to concede (change their preferred action) to reach an agreement.
\item If there are users whose willingness to accept changing their preferred action for the conflicting target user is low (Lines 8-14), then the mediator considers two cases: (i) if there are at least two users with low willingness and different preferred actions, then, according to concession rule \ref{IU}
, the action to be taken should be denying the conflicting target user access to the item in question; (ii) otherwise, rule \ref{IDM} applies so that the users that have high willingness will concede and the user/users who has/have low willingness will determine the action that is finally chosen as the solution. 

\end{itemize}

The complexity of Algorithm \ref{al:cr} is $\mathcal{O}(|C|\times |N|^2)$. That is, for each conflict, we need to know for each negotiating agent what is her willingness, which can be calculated in constant time as the sensitivity would only need to be calculated once for all conflicts, and the relative importance of each particular conflicting user can be obtained in constant time.  
Note that in the very worst case; i.e., all users are negotiating users and all users are at the same time conflicting, 
then the complexity of Algorithm \ref{al:cr} would be $\mathcal{O}(|U|^3)$.
\begin{algorithm}
\begin{small}
\begin{algorithmic}[1]
\Require $N$, $P_{n_1},\mathellipsis,P_{n_{\mid N\mid}}$, $C$
\Ensure $\vec{o}$
\ForAll{$c\in C$}
\State
\If {$\forall n\in N, \ \mathcal{W}(n,c) \text{ is HIGH}$}
\State $o[c]\gets \text{modified\_majority}(P_{n_1},\mathellipsis,P_{n_{\mid N\mid}},c)$ 
\State \textbf{continue}
\EndIf
\State
\If {$\exists a\in N, \ \mathcal{W}(a,c) \text{ is LOW }$}
\If {$\exists b\in N, \ \mathcal{W}(b,c) \text{ is LOW }\wedge v_a[c]\neq v_b[c]$}
\State $o[c]\gets 0$
\Else
\State $o[c]\gets v_a[c]$
\EndIf
\EndIf
\EndFor
\end{algorithmic}
\end{small}
\caption{Conflict Resolution}
\label{al:cr}
\end{algorithm}

\begin{table}[h]
\vspace{-0.25cm}
\begin{center}
\begin{tabular}{|c|c|c|} \hline
&Eve & Frank \\ \hline
Alice  & HIGH & LOW \\ \hline
Bob & LOW & HIGH \\ \hline
\end{tabular}
\caption{Fuzzy Memberships of willingness for Example 3.}
\label{tab:loss}
\end{center}
\end{table}
\begin{example}
Suppose again Example 1 and consider the willingness values calculated in Example 2. Table \ref{tab:loss} shows the fuzzy set membership for negotiating users Alice and Bob in case they would accept changing their most preferred action for the conflicting target users $C=\{\text{Eve},\text{Frank}\}$. We can see that for Alice and Eve IDM rule applies, so that the mediator assumes that Alice would concede (in this case, to accept granting Eve access to the item). As Bob has willingness LOW to change his preferred action for Eve, then the action suggested by this user would be taken to solve the conflict, and the computed solution would be to grant Eve access to the item. 
%
%
Regarding Frank, we have a similar situation. In this case, the willingness is HIGH for Bob, so that IDM rule applies and Bob would concede. As there is only one negotiating user (Alice) with willingness LOW, then the action suggested by this user is taken to solve the conflict. Therefore, the solution to the conflict would be to deny Frank access to the item. The resulting action vector for the item would be $\vec{o}=\{1,1,1,0\}$; i.e., Charlie, Dan and Eve would be granted access to the item while Frank would be denied access to the item.
\end{example}

\section{User Study}
\label{sec:evaluation}
The aim of this section is to compare the performance of our proposed mechanism to other existing approaches in terms of what users would do themselves manually in a number of situations. To this aim, we conducted the user study described below. 

\subsection{Method}
We sought to explore situations with different degrees of sensitivity, as users' behaviour to resolve conflicts may be different depending on how sensitive items are. However, this would have involved participants sharing with us sensitive items of them. Participants sharing sensitive information in user studies about privacy in Social Media was already identified as problematic in related literature \cite{wang2011regretted}, as participants would always seem reluctant to share sensitive information, which biases the study towards non-sensitive issues only. Indeed, this reluctance to share information that may be sensitive with researchers during user surveys is not only associated with studies about privacy and Social Media, but it has also been extensively proven to happen in many other survey situations, including other scientific disciplines such as psychology \cite{tourangeau2007sensitive}. A possible alternative to avoid this problem could be one in which participants just self-report how they behave when they experience a multi-party privacy conflict without asking for any sensitive information of them. However, the results obtained in that case may not match participants' actual behaviour in practice, as previous research on privacy and Social Media showed that there is a dichotomy between users' stated privacy attitudes and their actual behaviour \cite{acquisti2006imagined}. As a trade-off between these two alternatives, we chose to recreate \emph{situations} in which participants would be \emph{immersed}, following a similar approach to \cite{mancini2010contravision}, maximising actual behaviour elicitation while avoiding biasing the study to non-sensitive situations only. To this aim, we described a situation to the participants and asked them to immerse themselves in the situation by thinking they were a particular person in a particular photo that was to be shared through a Social Media site and that they were tagged in it, and participants showed very different individual privacy policies and concession decisions depending on the situation as detailed below. Each participant was presented with 10 different scenarios. Scenarios were different across participants as they were composed of: (i) one photo involving multiple users; and (ii) a conflict created based on the individual privacy policy the participant specified for the photo. As we had 50 participants (as detailed below), we were able to gather participant-specified data relative to 500 different scenarios. Photos referred to different situations (e.g., travelling, playing with friends, partying, dating, etc.) and were of different sensitivities \textit{a priori} --- though the participants were asked to specify their privacy policy for the photo as their first task for each scenario (as detailed below), which was different according to how sensitive each photo was for each participant. 

We developed a web application that presented the participants with the photos, stored the individual privacy policy they selected for each photo, generated conflicts, and stored whether or not participants would \emph{concede} during a negotiation in the scenarios presented. 
For each scenario, participants completed the following two tasks using the application:
\vspace{-0.15cm}
\begin{enumerate}
\itemsep0cm
\item \textbf{Definition of the Individual Privacy Policy}.  
Each participant was asked to define her/his most preferred privacy policy for each photo. 

\item \textbf{Conflict and Concession Question}. Once the participants defined their individual privacy policy for the photo, a conflict was generated. That is, we told the participants that one or more of the other people in the photo had a different most preferred action for one particular person, specifying the relationship type and strength the participant would have to this person. For instance, if the participant only wanted to share the photo with close friends, we told her/him that the other people in the photo wanted to share the photo with someone that was her/his acquaintance. Where multiple options were available to generate a conflict, we chose one of them randomly. Then, we asked participants whether or not they would concede and change their most preferred action for that person to solve the conflict with the other people depicted in the photo.

\end{enumerate}

\vspace{-8pt}

\subsection{Participants}
We recruited 50 participants via e-mail including university students, academic and non-academic staff, as well as other people not related to academia who volunteered to participate in the study. Participants completed the study online using the web application developed to that end (as detailed above). Before starting, the application showed the information to be gathered and participants needed to consent to continue. Table \ref{tab:dem} summarises participants' demographics (gender, age, job), Social Media use (number of accounts in different Social Media sites, and frequency of use), and if they were concerned about their privacy in Social Media (Priv. concerned). 

\begin{table}[h]
\begin{footnotesize}
\begin{center}
\begin{tabular}{|c|c|} \hline
\textbf{Variable} & \textbf{Distribution}\\ \hline
Gender & female (42\%), male (58\%) \\ \hline
Age & 18-24 (18\%), 25-30 (36\%), 31-40 (24\%), \\
& 41-50 (10\%), 51-60 (6\%), 60+(6\%)  \\ \hline
Job  & Agriculture(4\%), Arts(2\%), Computers(26\%),\\
&Design(6\%),Education(16\%),Engineering (10\%), \\
&Management(4\%),Media(2\%),\\
&Research(14\%),Sales(2\%), Other(14\%)\\ \hline
\# accounts & 0(4\%),1(30\%),2(18\%),3(8\%),4(12\%),4+(28\%)\\ \hline
Freq. of use & monthly- (18\%), monthly (10\%), \\
& weekly(10\%), daily (26\%), daily+(36\%) \\ \hline
Priv. concerned& not much(36\%), yes(26\%), very much(36\%) \\ \hline
\end{tabular}
\vspace{0.05cm}
\caption{Participants' demographics, Social Media use, and privacy concern.}
\label{tab:dem}
\end{center}
\end{footnotesize}
\vspace{-1cm}
\end{table}

\subsection{Results}
\label{ar-results} 
The results gathered through the web application were compared to the results that would have been obtained if our proposed mechanism was 
applied to the scenarios and if state-of-the-art automated voting mechanisms were applied. To this aim, we looked at the privacy policy defined by the participant and the conflict generated by the application for each situation. This determined participants' most preferred action for the conflict (to be considered by our proposed mechanism and state-of-the-art voting mechanisms), as well as the willingness to change it (used to determine the concession rule our mechanism would apply in each case). 
In particular, we compared the results that would have been obtained applying our proposed mechanism to those that would have been obtained applying the general voting mechanisms used in state-of-the-art automated approaches: 
\begin{itemize}
\itemsep0cm
\item \textbf{Uploader overwrites} (UO), the conflict is solved selecting the action preferred by the user that uploads the item. This is the strategy currently followed by most Social Media Sites (Facebook, etc.).
\item \textbf{Majority voting} (MV) \cite{carminati2011collaborative}, the conflict is solved selecting the action most preferred by the majority of the negotiating users. 
\item \textbf{Veto voting} (VV) \cite{thomas2010unfriendly}, if there is one negotiating user whose most preferred action is denying access, the conflict is solved by denying access to the item.
\end{itemize}  

\begin{figure}[!t]
\centering
\includegraphics[scale=0.32]{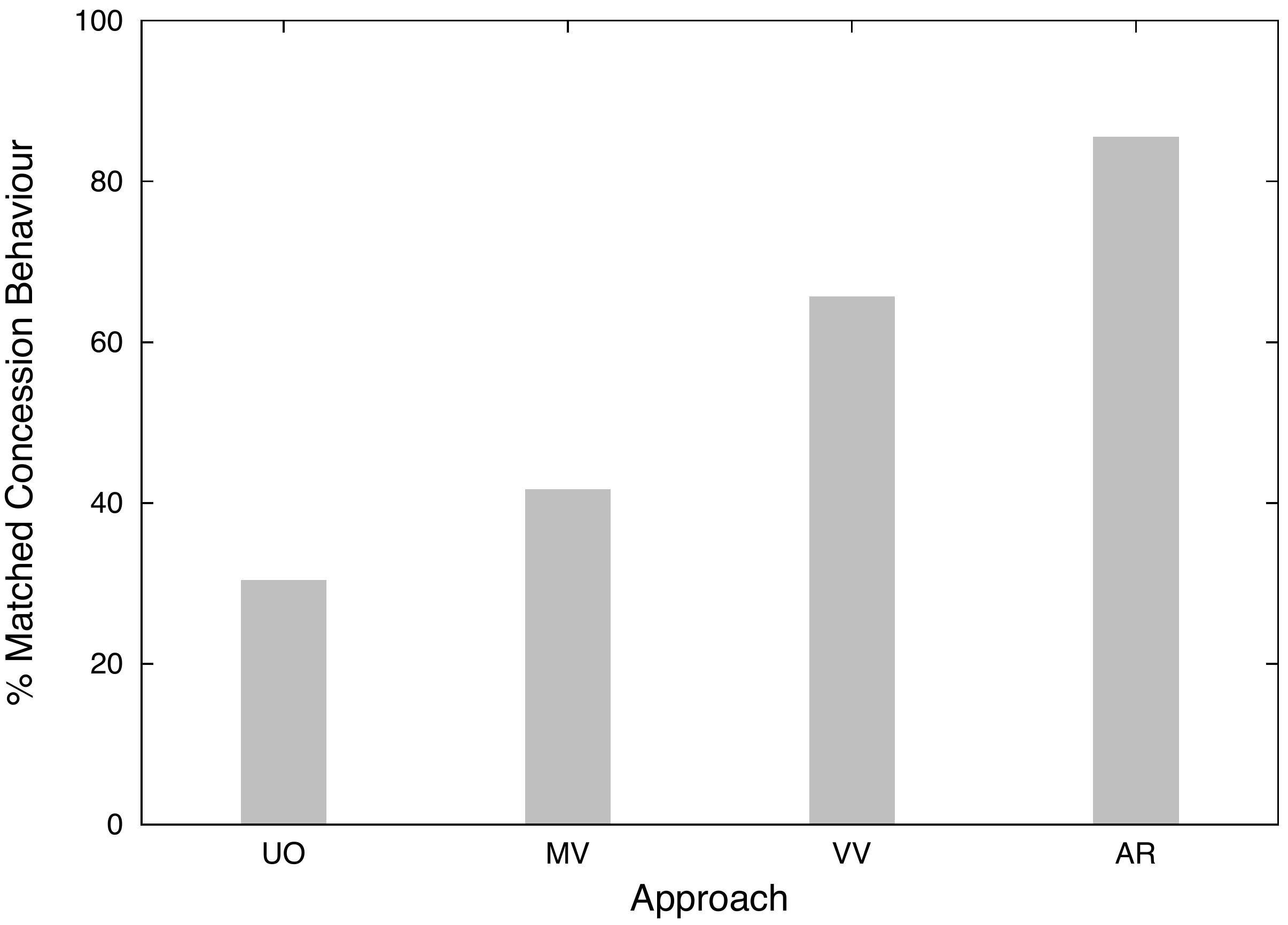}
\vspace{-0.35cm}
\caption{Percentage of times each approach matched concession behaviour.}
\label{fig:comparison}
\vspace{-0.4cm}
\end{figure}

Figure \ref{fig:comparison} shows the results for each of the above voting mechanisms as well as the results for our proposed mechanism for automated conflict resolution (labelled AR in the figure). In particular, it shows the percentage of times each mechanism matched participants' concession behaviour in the scenarios above. We can observe that our proposed mechanism AR clearly outperformed UO, MV, and VV. This is because these mechanisms lack enough flexibility to model actual user behaviour across different situations in this domain, as they only consider the most preferred action for each negotiating user as a vote without considering the particular situation. 
We can also observe that UO is very far from what users did themselves, which is mainly due to UO not being collaborative at all ---i.e., the preferences of the other parties are not considered. MV performs a bit better than UO, but it is still far from what participants did themselves. This is mostly due to the situations in which even if the majority of users would like to share an item in the first instance, they could reconsider this if there is/are one/multiple user/(s) that would prefer not sharing because this could have privacy consequences for them. 

We can also see in Figure \ref{fig:comparison} that VV performs better than UO and MV. This result confirms that negotiating users are many times open to accept not sharing an item if this can cause privacy breaches to one of them --- as also modelled in our proposed mechanism AR. However, VV is too restrictive to be suitable for all situations. This is because there are also situations in which the user/s whose most preferred action is denying access may not mind granting access due to many reasons. In these cases, VV would suggest solutions that mean losing sharing opportunities. For instance, as stated earlier, Alice and Bob could be depicted in a photo with very low sensitivity --- e.g., a photo in which both Alice and Bob are depicted with a view to a monument --- and both of them could have defined privacy policies for the photo so that all their friends can see it. Suppose that Charlie is friend of Alice but is distant acquaintance of Bob, so according to Alice's privacy policy Charlie should be granted access to the photo but according to Bob's privacy policy Charlie should not be granted access to the item. However, given that the photo is not sensitive for Bob, Bob would probably accept sharing also with Charlie. VV would not consider this concession, and the solution to solve the conflict would be not sharing with Charlie, so it would be a lost sharing opportunity and Alice may not even accept the solution. 
In contrast, our mechanism is able to adapt to the particular situation, being as restrictive as VV if needed but also considering the cases in which \emph{concessions} about granting access are to happen ---as the example above, in which the I do not mind (IDM) rule would have picked that Bob would concede, so that the final solution would be to share with Charlie (recall the item was not sensitive to Bob).

\begin{figure}[!t]
\centering
\includegraphics[scale=0.32]{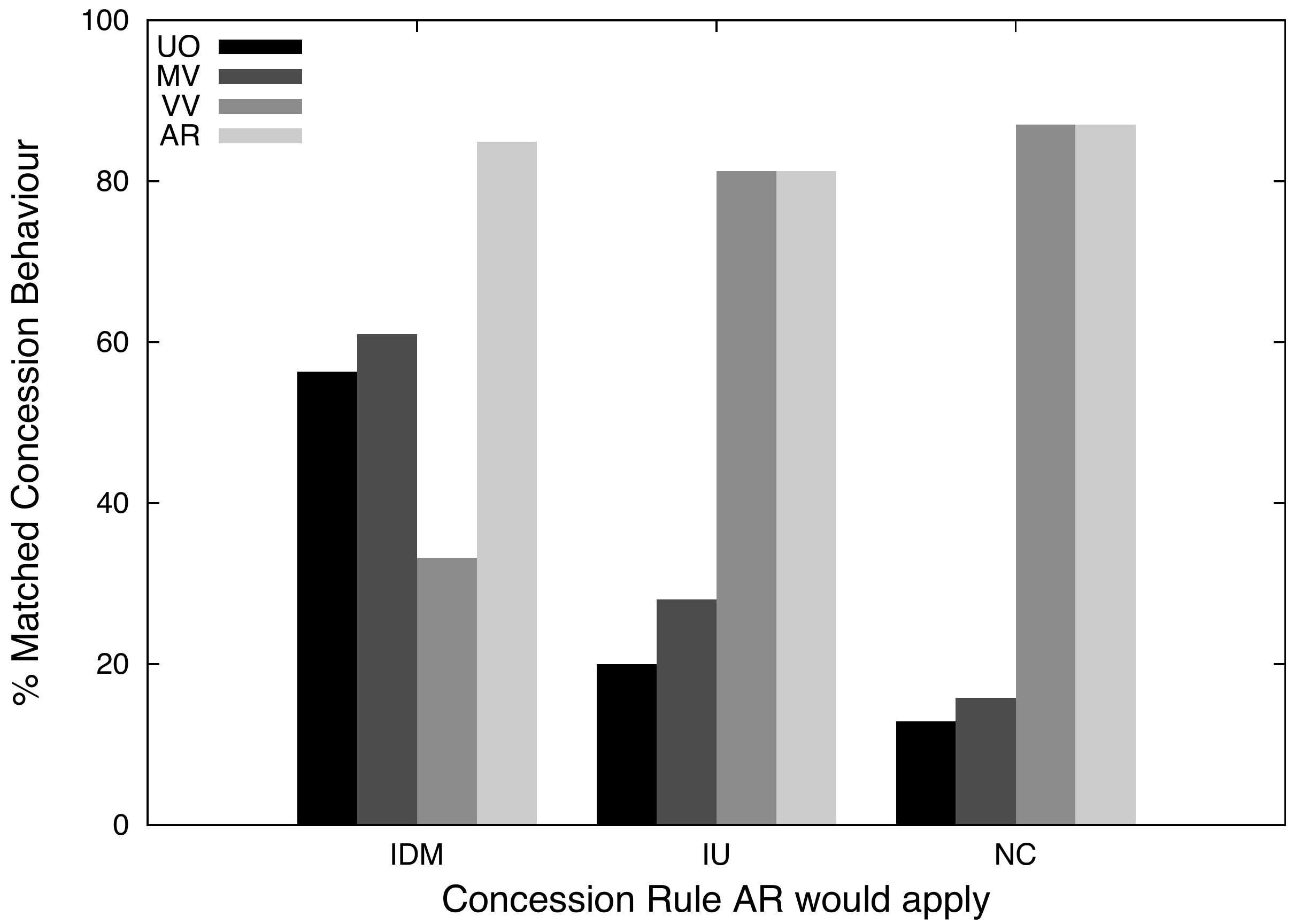}
\vspace{-0.35cm}
\caption{Percentage of times each approach matched concession behaviour broken down by the concession rule AR would apply (IDM - I do not mind, IU - I understand, NC - No concession).}
\label{fig:comparisonxrule}
\vspace{-0.4cm}
\end{figure}

We also sought to explore more closely how each of the concession rules in our proposed mechanism contributed to its performance as well as how state-of-the-art voting mechanisms would work in each case. Table \ref{tab:results} shows for each concession rule the number of times that each rule would have been applied (\# Instantiations) in the 500 situations and Figure \ref{fig:comparisonxrule} shows the performance of each approach broken down by the concession rule that would have been applied for each situation. We can observe that performance was similar across concession rules for our proposed mechanism AR; i.e., once a particular concession rule instantiated for a situation, it usually matched users' behaviour with respect to concessions. In particular, we observe that the three concession rules in our mechanism obtain better results than the state-of-art approaches. We can also observe that the performance of state-of-the-art voting mechanisms significantly varied according to the concession rule AR would apply. This confirms the fact that static ways of aggregating preferences (as those used in state-of-the-art voting mechanisms) are not desirable in this domain, because the concessions that may happen to resolve multi-party privacy conflicts clearly depend on the particular situation ---as captured by the variables considered by AR's concession rules; i.e., individual privacy preferences of each user, the sensitivity of the item to be shared, and the relative importance of the conflicting target user. 

\begin{table}[!h]
\begin{footnotesize}
\begin{center}
\begin{tabular}{|c|c|} \hline
Concession Rule &\# Instantiations \\ \hline
I do not mind (IDM) & 172 \\ \hline
I understand (IU)  & 111 \\ \hline
No concession (NC)  & 217 \\ \hline
\hline
\textbf{Total} & 500   \\ \hline
\end{tabular}
\vspace{0.025cm}
\caption{Number of times each AR concession rule would have been applied.}
\label{tab:results}
\end{center}
\end{footnotesize}
\vspace{-0.5cm}
\end{table}

Finally, we sought to find any correlation that could exist between participants' data ---demographics, social media use and privacy concern--- and whether or not participants behaved according to the concession rule instantiated for each situation. 
To this aim, we calculated the information gain (IG) --- i.e., the reduction in entropy --- that each variable produced on whether the participant followed the corresponding rule or not once it was instantiated, and the Pearson's correlation coefficient (CC)
. Table \ref{tab:correlations} summarises the values for each rule. IGs and CCs were negligible and not statistically significant (i.e., $p>0.05$). Thus, users' characteristics like the privacy concern, age, gender, profession, studies, and social media use did not have any significant effect on whether participants followed a concession rule once the rule was instantiated for a particular situation. Note, however, the particular concession rule instantiated in each situation depended on the individual privacy policy of each user, the sensitivity of the item for the user, and the relative importance of the conflicting target user as stated above, which may vary from participant to participant. The important thing is that once a rule was instantiated, the variables above did not influence whether the particular instantiated rule was successful in matching user behaviour or not. 
In other words, users' characteristics (e.g., demographics, privacy concern, etc.) may determine the individual privacy policies users choose, which in turn determine the rules that are instantiated for a given situation; but users' characteristics do not determine whether users' concession behaviour matches that of the rule instantiated. This suggests the mechanism proposed in this paper captures general user behaviour and would be able to adapt to both different situations and users.



\begin{table}[!t]
\begin{footnotesize}
\begin{center}
\begin{tabular}{c|c|c|c|c|c|c|} \cline{2-7}
 & \multicolumn{2}{|c|}{Rule 1} & \multicolumn{2}{|c|}{Rule 2}  & \multicolumn{2}{|c|}{Rule 3} \\ \cline{2-7} 
 & IG & CC  & IG & CC & IG & CC \\ \hline
\multicolumn{1}{|r|}{Age} & 0 &  0.04 & 0 & 0 & 0 & -0.10\\ \hline
\multicolumn{1}{|r|}{Gender}  & 0.06 &  0.14 & 0.06 & 0.13  & 0 & -0.06 \\ \hline
\multicolumn{1}{|r|}{Job}  & 0.08 & -0.18 & 0.08 & -0.17 & 0.04 & 0.11 \\ \hline 
\multicolumn{1}{|r|}{Studies} & 0 & 0.18 & 0 & 0.16 & 0 & 0.017 \\ \hline 
\multicolumn{1}{|r|}{Freq. of use}  & 0.03 & 0.05 & 0.06 & 0.08 &  0.02 &  -0.07 \\ \hline
\multicolumn{1}{|r|}{\# Accounts} & 0 & 0.14 & 0 & 0.16 & 0 & -0.05 \\ \hline
\multicolumn{1}{|r|}{Priv. Concern} & 0 & 0.13 & 0 & 0.16 & 0 & 0.04 \\ \hline
\end{tabular}
\vspace{0.1cm}
\caption{IGs and CCs for each rule based on participants' demographics, social media use, and privacy concern. 
}
\label{tab:correlations}
\end{center}
\end{footnotesize}
\vspace{-0.7cm}
\end{table}

\section{Discussion}
%

The results of the user study suggest that our mechanism was able to match participants’ concession behaviour significantly more often than other existing approaches. The results also showed the benefits that an
adaptive mechanism like the one we presented in this paper can provide with respect to more static ways of aggregating users’ individual privacy preferences, which are unable to adapt to different situations and were far from what the users did themselves.
Importantly, our mechanism is agnostic to and independent from how a user interface \emph{communicates} the suggested solutions to users and gets \emph{feedback} from them. First, privacy visualisation tools already proved to be highly usable for social media could be used to show and/or modify the suggested solution, such as AudienceView \cite{lipford2008understanding}, PViz \cite{mazzia2012pviz}, or the Expandable Grid \cite{reeder2008expandable}. Second, users could define a default response to the solutions suggested, e.g., \emph{always accept the suggested solution without asking me}\footnote{This would be similar to the tagging mechanism of Facebook, which users can configure to be notified for confirmation about tags before they become active or to just go ahead without confirmation.}, which, as shown in the evaluation (Section \ref{sec:evaluation}), would actually match user behaviour very accurately. Other \emph{suitable defaults} could be applied based on approaches like \cite{toch2010generating,watson2015mapping,hirschprung2015simplifying}, or users' responses could be (semi-)automated based on the concession rules instantiated in each situation, using any of the machine-learning approaches shown to work very well in social media privacy settings \cite{fang2010privacy,squicciarini2011a3p}.

We considered the individual privacy preferences of each individual involved in an item, sensitivity of the item and the relative importance of the target to determine a user's willingness to concede when a multi-party privacy conflict arises. Although accuracy results presented in the previous section are encouraging, this does not mean that there are no other factors that play a role to determine concessions. For instance, in e-commerce domains the strength of relationships among negotiators themselves is also known to influence to what extent negotiators are willing to concede during a negotiation \cite{sierra07}. Future research should look into how other factors could help further increase the accuracy of the mechanism presented here.

Finally, we focused on detecting and resolving conflicts once we know the parties that co-own an item and have their individual privacy policies for the item. However, we are not proposing a method to automatically detect which items are co-owned and by whom they are co-owned. This is a different problem that is out of the scope of this paper. For example, Facebook researchers developed a face recognition method that correctly identifies Facebook users in 97.35\% of the times \cite{taigman2014deepface}. Also, it could be the case that a person does not have an account in a given social media. In that case, her face could be \emph{preventively} blurred \cite{ilia2015face}.  Blurring faces may seriously diminish the \emph{utility} of sharing information in social media, but it could also be a good alternative if no agreement is reached between negotiators to ensure an individual (not collective) privacy baseline is achieved.

\section{Related Work}
\label{sec:relwork}
Until now, very few researchers considered the problem of resolving conflicts in multi-party privacy management for Social Media. Wishart et al.\ \cite{wishart2010collaborative} proposed a method to define privacy policies collaboratively. In their approach all of the parties involved can define strong and weak privacy preferences. 
However, this approach does not involve any automated method to solve conflicts, only some suggestions that the users might want to consider when they try to solve the conflicts manually.

The work described in \cite{squicciarini2009collective} 
is based on an incentive mechanism where users are rewarded with a quantity of numeraire each time they share information or acknowledge the presence of other users (called co-owners) who are affected by the same item
. When there are conflicts 
among co-owners' policies, users can spend their numeraire bidding for the policy that is best for them. Then, the use of the Clark Tax mechanism is suggested to obtain the highest bid. As stated in \cite{hu2012detecting}, 
users may have difficulties to comprehend the mechanism and specify appropriate bid values in auctions. 
Furthermore, users that earned much numeraire in the past will have more numeraire to spend it at will, potentially leading to unilateral decisions.

In \cite{hu2012detecting} 
users must manually define for each item: the privacy settings for the item, their trust to the other users, the sensitivity of the item, and how much privacy risk they would like to take. These parameters are used to calculate what the authors call privacy risk and sharing loss on segments --- they define segments as the set of conflicting target users among a set of negotiating users. Then, based on these measures all of the conflicting target users in each segment are assigned the same action. That is, all of the conflicts that a set of negotiating users have would be solved either by granting or denying access. Clearly, not considering that each individual conflict can have a different solution leads to outcomes that are far from what the users would be willing to accept. Moreover, due to how the privacy risk and sharing loss metrics are defined, solutions are likely to be the actions preferred by the majority of negotiating users, which can be many times far from the actual behaviour of users as shown in Section \ref{sec:evaluation}. 

There are also related approaches based on voting in the literature \cite{carminati2011collaborative,thomas2010unfriendly}. In these cases, a third party collects the decision to be taken (granting/denying) for a particular friend from each party. Then, the authors propose to aggregate a final decision based on one of the voting rules already been described in Section \ref{sec:evaluation} --- i.e., uploader overwrites (UO), majority voting (MV), and veto voting (VV). These approaches are static, in the sense that they always aggregate individual votes in the same way by following the same voting rule. Thus, these approaches are unable to adapt to different situations that can motivate different concessions by the negotiating users, which makes these approaches unable to match the actual behaviour of users many times, as shown in Section \ref{sec:evaluation}. Only in \cite{hu2012multiparty}, the authors consider that 
a different voting rule could be applied depending on the situation. However, it is the user who uploads/posts the item the one who chooses \emph{manually} which one of the voting rules (UO,MV,VV) to apply for each item. 
The main problem with this --- apart from having to specify the voting rule manually for every item --- is that the choice of the voting rule to be applied is unilateral. That is, the user that uploads the item decides the rule to apply without considering the rest of the negotiating users' preferences, which becomes a unilateral decision on a multi-party setting. 
Moreover, it might actually be quite difficult for the user that uploads the item to anticipate which voting rule would produce the best result without knowing the preferences of the other users. 

Finally, the problem of negotiating a solution to multi-party conflicts, has also been recently analysed from a game-theoretic point of view \cite{hu2014game,such2014privacy}. These proposals provide an elegant analytic framework proposing negotiation protocols to study the problem and the solutions that can be obtained using well-known game-theoretic solution concepts such as the Nash equilibrium. However, as shown in \cite{hu2014game}, these proposals may not always work well in practice, as they do not capture the social idiosyncrasies considered by users in the real life when they face multi-party privacy conflicts, and users' behaviour is far from perfectly rational as assumed in these game-theoretic approaches --- e.g., refer to \cite{lampinen2011we,wisniewski2012fighting}.

\section{Conclusions}
\label{sec:conclusions}

In this paper, we present the first mechanism for detecting and resolving privacy conflicts in Social Media that is based on current empirical evidence about privacy negotiations and disclosure driving factors in Social Media and is able to adapt the conflict resolution strategy based on the particular situation. In a nutshell, 
the mediator firstly inspects the individual privacy policies of all users involved looking for possible conflicts. If conflicts are found, the mediator proposes a solution for each conflict according to a set of concession rules that model how users would actually negotiate in this domain. 

We conducted a user study comparing our mechanism to what users would do themselves in a number of situations. The results obtained suggest that our mechanism was able to match participants' concession behaviour significantly more often than other existing approaches. This has 
the potential to reduce the amount of \emph{manual} user interventions to achieve a satisfactory solution for all parties involved in multi-party privacy conflicts.
Moreover, the study also showed the benefits that an adaptive mechanism like the one we presented in this paper can provide with respect to more static ways of aggregating users' individual privacy preferences, which are unable to adapt to different situations and were far from what the users did themselves. 

The research presented in this paper is a stepping stone towards more automated resolution of conflicts in multi-party privacy management for Social Media. 
As future work, we plan to continue researching on what makes users concede or not when solving conflicts in this domain. In particular, we are also interested in exploring if there are other factors that could also play a role in this, like for instance if concessions may be influenced by previous negotiations with the same negotiating users or the relationships between negotiators themselves. 

\bibliographystyle{IEEEtran}
\bibliography{IEEEabrv,policy-negotiation}

\appendices

\section{Proof of the Principles}
\label{sec:proofs}
\begin{principle}
Content should not be shared if it is detrimental to one of the users involved
\end{principle}
\begin{proof}
We prove that solutions proposed by the conflict resolution algorithm (Algorithm \ref{al:cr}) follow this principle by contradiction. Suppose that Principle 1 does not hold, thus given a negotiating user $a\in N$ who does not want to share an item with a conflicting target user $c\in C$ (i.e, $v_a[c]=0$) and doing this is detrimental to her (i.e., $\mathcal{W}(a,c)$ is $low$)
, the solution to the conflict will not respect the decision made by $a$ and it will be sharing the item (i.e., $o[c]=1$). However, Algorithm \ref{al:cr} can only output value $1$ for $o[c]$ in two cases:
\begin{enumerate}
	\item As a result of a modified majority (line 4). However, this is only executed when there is not a user $u\in N$ such that $\mathcal{W}(u,c)$ is $low$,  which contradicts our assumption.
	\item As a direct assignation from a user who prefers to share the item (line 12). However, this is only executed when there is not another user $u\in N$ such that $v_u[c]=0$ and $\mathcal{W}(u,c)$ is $low$ which contradicts our assumption.
	\end{enumerate}
	\vspace{-10pt}
\end{proof}

\begin{principle}
If an item is not detrimental to any of the users involved and there is any user for whom sharing is important, the item should be shared.
\end{principle}
\begin{proof}
We prove that solutions proposed by the conflict resolution algorithm (Algorithm \ref{al:cr}) follow this principle by contradiction. Suppose that Principle 2 does not hold, thus given a negotiating user $a\in N$ who wants to share an item with a conflicting target user $c\in C$ (i.e, $v_a[c]=1$) because doing this is important to her (i.e., $\mathcal{W}(a,c)$ is $low$), and that there does not exist a negotiating user $b\in N$ who does not want to share the item with $c$ (i.e., $v_b[c]=0$) and the item is detrimental to her (i.e., $\mathcal{W}(b,c)$ is $low$), the solution to the conflict will not respect the decision made by $a$ and it will be not sharing the item (i.e., $o[c]=0$). However, Algorithm \ref{al:cr} can only output value $0$ for $o[c]$ in three cases:
\begin{enumerate}
	\item As a result of a modified majority (line 4). However, this is only executed when there is not a user $u\in N$ such that $\mathcal{W}(u,c)$ is $low$,  which contradicts our assumption.
	\item As a direct assignation of not sharing (line 10). However, this is only executed when there is another user $u\in N$ such that $v_u[c]=0$ and $\mathcal{W}(u,c)$ is $low$ which contradicts our assumption.
	\item As a direct assignation from a user who prefers not to share the item (line 12). However, this would only be executed when there exists another user $u\in N$ such that $v_u[c]=0$ and $\mathcal{W}(u,c)$ is $low$ which contradicts our assumption.
	\end{enumerate}
		\vspace{-10pt}
\end{proof}

Finally, the proof for Principle 3 is omitted because of lack of space, but it is trivial to prove that for all other cases not considered in Principles 1 and 2, the modified majority voting will aggregate all users' preferences. 

\end{document}